\documentclass[default,iicol,sn-mathphys-num]{sn-jnl}

\usepackage{graphicx}%
\usepackage{multirow}%
\usepackage{amsmath,amssymb,amsfonts}%
\usepackage{amsthm}%
\usepackage{mathrsfs}%
\usepackage[title]{appendix}%
\usepackage{xcolor}%
\usepackage{textcomp}%
\usepackage{manyfoot}%
\usepackage{booktabs}%
\usepackage{algorithm}%
\usepackage{algorithmicx}%
\usepackage{algpseudocode}%
\usepackage{listings}%
\usepackage{caption}%

\theoremstyle{thmstyleone}%

\theoremstyle{thmstyletwo}%

\theoremstyle{thmstylethree}%

\begin{document}

\title[Article Title]{Interlayer coupling enhanced superconductivity near 100 K in La$_{3-x}$Nd$_x$Ni$_2$O$_7$}

\author[1]{\fnm{Zhengyang} \sur{Qiu}}
\equalcont{These authors contributed equally to this work.}
\author[1]{\fnm{Junfeng} \sur{Chen}}
\equalcont{These authors contributed equally to this work.}
\author[2]{\fnm{Dmitrii V.} \sur{Semenok}}
\equalcont{These authors contributed equally to this work.}
\author[3]{\fnm{Qingyi} \sur{Zhong}}
\author[2]{\fnm{Di} \sur{Zhou}}
\author[1]{\fnm{Jingyuan} \sur{Li}}
\author[1]{\fnm{Peiyue} \sur{Ma}}
\author[1]{\fnm{Xing} \sur{Huang}}
\author[1]{\fnm{Mengwu} \sur{Huo}}
\author[1]{\fnm{Tao} \sur{Xie}}
\author[1]{\fnm{Xiang} \sur{Chen}}
\author[4,5]{\fnm{Ho-kwang} \sur{Mao}}
\author*[4,5]{\fnm{Viktor} \sur{Struzhkin}}\email{viktor.struzhkin@hpstar.ac.cn}
\author*[3]{\fnm{Hualei} \sur{Sun}}\email{sunhlei@mail.sysu.edu.cn}
\author*[1]{\fnm{Meng} \sur{Wang}}\email{wangmeng5@mail.sysu.edu.cn}

\affil[1]{\orgdiv{Center for Neutron Science and Technology at School of Physics, Guangdong Provincial Key Laboratory of Magnetoelectric Physics and Devices}, \orgname{Sun Yat-Sen University}, \city{Guangzhou}, \postcode{510275}, \state{Guangdong}, \country{China}}
\affil[2]{\orgname{Center for High Pressure Science and Technology Advanced Research (HPSTAR)}, \city{Beijing}, \postcode{100193}, \country{China}}
\affil[3]{\orgdiv{School of Science}, \orgname{Sun Yat-Sen University}, \city{Shenzhen}, \postcode{518107}, \state{Guangdong}, \country{China}}
\affil[4]{\orgname{Center for High Pressure Science and Technology Advanced Research (HPSTAR)}, \city{Shanghai}, \postcode{201203}, \country{China}}
\affil[5]{\orgname{Shanghai Key Laboratory of Material Frontiers Research in Extreme Environments (MFree), Shanghai Advanced Research in Physical Sciences (SHARPS)}, \orgaddress{\street{Pudong}}, \city{Shanghai}, \postcode{201203}, \country{China}}
\abstract{Systematically controlling the superconducting transition temperature ($T_\text{c}$) in the bilayer Ruddlesden-Popper nickelate La$_3$Ni$_2$O$_7$ remains a significant challenge. Here, we address this by synthesizing high-quality polycrystalline La$_{3-x}$Nd$_x$Ni$_2$O$_7$ ($0 \leq x \leq 2.4$) with record-level rare-earth substitution. Nd doping compresses the lattice, particularly along the $c$ axis, enhances the spin density wave transition temperature, and elevates the pressure required for the orthorhombic-to-tetragonal structural transition. Superconductivity is observed across all doping levels under high pressures, with the onset $T_\text{c}$ rising to $\sim$93~K for $x = 2.1$ and $2.4$ from the electronic transport measurement. Using the radio-frequency transmission technique, newly applied to nickelate superconductors, we detect signatures of superconductivity at $98 \pm 2$~K in the $x=2.4$ compound, pushing the $T_\text{c}$ frontier further. We identify a universal linear relationship where $T_\text{c}$ decreases with the $c$-axis lattice parameter at a rate of approximately $-28$~K/\AA, demonstrating that enhanced interlayer magnetic exchange coupling is the dominant mechanism for superconducting pairing. Our work establishes the critical role of magnetism and provides a unified structural descriptor for elevating $T_\text{c}$ in bilayer nickelates.
}
\keywords{High temperature superconductivity, nickelate superconductor, high pressure}

\maketitle

\section*{Introduction}  

The discovery of superconductivity in square-planar nickelates, specifically the infinite-layer compounds ($R_{n+1}$Ni$_n$O$_{2n+2}$)\cite{Li2019,Zeng2022,Lee2023,Parzyck2025,Sahib2025,Pan2022,Chow2025b}, where $R$ is a rare-earth metal) and the Ruddlesden-Popper (RP) series ($R_{n+1}$Ni$_n$O$_{3n+1}$)\cite{Sun2023,Zhang2024h,Hou2023,Wang2024b,Zhang2024eff,Li2025i,Liu2025evi,Wen2024,Shi2025p,Zhou2025i,Ueki2025,Sakakibara2024t,Zhu2024s,Shi2025a,Xu2025c,Li2024sign,Zhang2025sc,Li2025si,Li2024str,Huang2024s,Pei2024,Zhang2025b,Chen2025l,Li2024de,Shi2025s,Xu2024,Feng2024,Wang2025c,Li2025a,Liu2025em}, has established a new family of high-temperature unconventional superconductors. The square-planar nickelates, long sought after due to their electronic similarities to cuprate superconductors\cite{Anisimov1999}, provide a compelling comparative system. In contrast, the RP nickelates exhibit distinct characteristics that set them apart from both cuprates and iron-based superconductors\cite{Keimer2015,Wang2024nor,Wang2025r}. Among the superconducting RP nickelates, which include La$_3$Ni$_2$O$_7$\cite{Sun2023,Li2025i,Hou2023,Zhang2024h,Zhang2024eff,Liu2025evi,Wen2024,Shi2025p,Zhou2025i,Ueki2025,Sakakibara2024t}, La$_4$Ni$_3$O$_{10}$\cite{Zhu2024s,Shi2025a,Xu2025c,Li2024sign,Sakakibara2024t,Zhang2025sc,Li2025si,Li2024str}, Pr$_4$Ni$_3$O$_{10}$\cite{Huang2024s,Pei2024,Zhang2025b,Chen2025l}, the hybrid La$_5$Ni$_3$O$_{11}$\cite{Li2024de,Shi2025s}, and their doped derivatives\cite{Wang2024b,Xu2024,Feng2024,Wang2025c,Li2025a,Liu2025em}, the pressurized bilayer RP phase (in its Sm-doped variants) holds the record with an onset $T_\text{c}$ of 92 K\cite{Li2025a,Zhong2025e}. Thin-film samples of bilayer RP nickelates at ambient pressure exhibit a lower $T_\text{c}$ of approximately $40-50$ K\cite{Ko2025,Zhou2025a,Liu2025s,Hao2025}. The mechanism underlying this significant variation in superconducting transition temperatures remains a subject of active investigation. However, systematic studies of the effects of doping and oxygen stoichiometry have been challenging, primarily due to the difficulty in synthesizing the pure bilayer RP phase\cite{Wang2024b,Feng2024,Li2025a}.

Here, we report the systematic synthesis of high-quality polycrystalline La$_{3-x}$Nd$_x$Ni$_2$O$_7$ ($0 \leq x \leq 2.4$) samples with record-high Nd substitution, achieved via a sol-gel method\cite{Zhang1994,Wang2024p}. Through comprehensive structural, electronic transport, and radio-frequency (RF) measurements under high pressure, we observe a maximum $T_\text{c}$ of 93 K in La$_{0.9}$Nd$_{2.1}$Ni$_2$O$_7$ from the electronic transport measurements, and a signature of superconductivity near 100 K in La$_{0.6}$Nd$_{2.4}$Ni$_2$O$_7$ from the RF response. The pressure required for the emergence of superconductivity increases significantly with Nd doping, revealing a direct connection between chemical-pressure-induced structural distortion and superconductivity. Compounds with high Nd-doping levels ($x = 2.1$--$2.4$) and the Sm-doped analogue ($x=1.5$), which possess comparable $c$ lattice constants, exhibit similar $T_\text{c}$. A comprehensive comparison reveals a linear relationship between the $c$ lattice constant and $T_\text{c}$ that holds for both pressurized bulk samples and thin films at ambient pressure. Our results thus identify the interlayer magnetic exchange coupling as a key intrinsic factor regulating superconductivity\cite{luo2023b,yang2024o,liu2024e,qu2024,Lu2024i1,Sakakibara2024p,Zhang2024str,Jiang2025t,Yang2023i,Lechermann2023}, offering a feasible pathway to enhance $T_\text{c}$ in RP nickelates.

\begin{figure*}[t]
\centering
  \includegraphics[width=1.8\columnwidth]{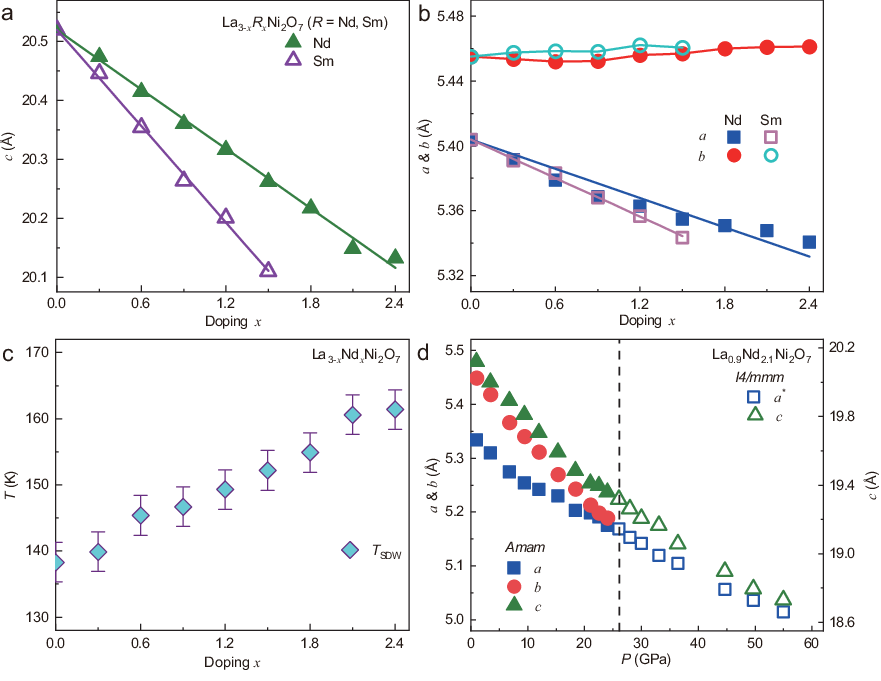}
    \caption{\textbf{ Structural characterizations of La$_{3-x}$Nd$_x$Ni$_2$O$_7$ under pressure.} \textbf{a,b,} Ambient-pressure lattice parameters of La$_{3-x}$Nd$_x$Ni$_2$O$_7$ as a function of doping level $x$, showing the $c$-axis parameter (\textbf{a}) and the $a$- and $b$-axis parameters (\textbf{b}) obtained from XRD refinements. For comparison, lattice parameters of La$_{3-x}$Sm$_x$Ni$_2$O$_7$ are included, adopted from Ref.\cite{Zhong2025e}. \textbf{c,} Doping dependence of the spin density wave transition. \textbf{d,} Pressure evolution of the lattice parameters $a$, $b$, and $c$ for La$_{0.9}$Nd$_{2.1}$Ni$_2$O$_7$ refined from synchrotron XRD data. Solid symbols represent the orthorhombic $Amam$ phase, while hollow symbols correspond to the tetragonal $I4/mmm$ phase.
    }
        \label{fig1}
\end{figure*}

\section*{Results and discussion}
\subsection*{Ambient and high-pressure structural characterizations}

We synthesized a series of doped La$_{3-x}$Nd$_x$Ni$_2$O$_7$ ($0 \leq x \leq 2.4$) polycrystalline samples using an optimized sol-gel method\cite{Zhang1994,Wang2024p}. X-ray diffraction (XRD) measurements confirm the retention of a pure bilayer phase (Extended Data Fig. 1). Figures \ref{fig1}a,b display the lattice constants for the Nd-doped and Sm-doped compounds grown under identical conditions. All compounds adopt the orthorhombic structure (space group $Amam$), consistent with pristine La$_3$Ni$_2$O$_7$ at ambient pressure. With increasing Nd or Sm doping concentration, the lattice parameters $a$ and $c$ undergo significant contraction, whereas $b$ remains largely unchanged\cite{Zhong2025e}. The ionic radius of Nd$^{3+}$ (0.995 \AA) is smaller than that of La$^{3+}$ (1.106 \AA) but larger than that of Sm$^{3+}$ (0.964 \AA). The compression of the $c$ lattice constant is less pronounced in Nd-doped compounds than in their Sm-doped counterparts. The refined lattice parameters for La$_{0.6}$Nd$_{2.4}$Ni$_2$O$_7$ at ambient pressure ($a = 5.341$ \AA, $b = 5.461$ \AA, $c = 20.132$ \AA) are remarkably similar to those of La$_{1.5}$Sm$_{1.5}$Ni$_2$O$_7$, suggesting that the doping level may approach the upper limit of thermodynamic stability achievable with the sol-gel method. The anisotropic compression reveals a more significant structural distortion along the $c$-axis compared to the $ab$ plane. As the Nd doping increases, the obtained compound becomes more insulating. The magnetic susceptibility is systematically enhanced (Extended Data Fig. 2). The kinks in the resistance and its derivative against temperature reveal that the spin density wave transition temperature evolves from 138 to 161 K (Fig. \ref{fig1}c), suggesting enhancement of the interlayer magnetic exchange interaction\cite{Chen2024ele,Xie2024s}. 

Synchrotron XRD was employed to study the structural evolution of La$_{3-x}$Nd$_{x}$Ni$_2$O$_7$ ($x=0.9$ and 2.1) under high pressure. The diffraction patterns are well indexed by the orthorhombic $Amam$ space group at low pressures and by the tetragonal $I4/mmm$ space group above a certain transition pressure. The structural transition pressure is identified by the merging of the (0 2 0) and (2 0 0) diffraction peaks (Extended Data Fig. 4). Nd doping is found to enhance the interapical oxygen distortion, which consequently increases the structural transition pressure from 13 GPa in the $x=0.9$ compound to 26 GPa in the $x=2.1$ compound, as shown in Fig. \ref{fig1}d and Extended Data Fig. 4.

\begin{figure*}[t]
\centering
  \includegraphics[width=1.8\columnwidth]{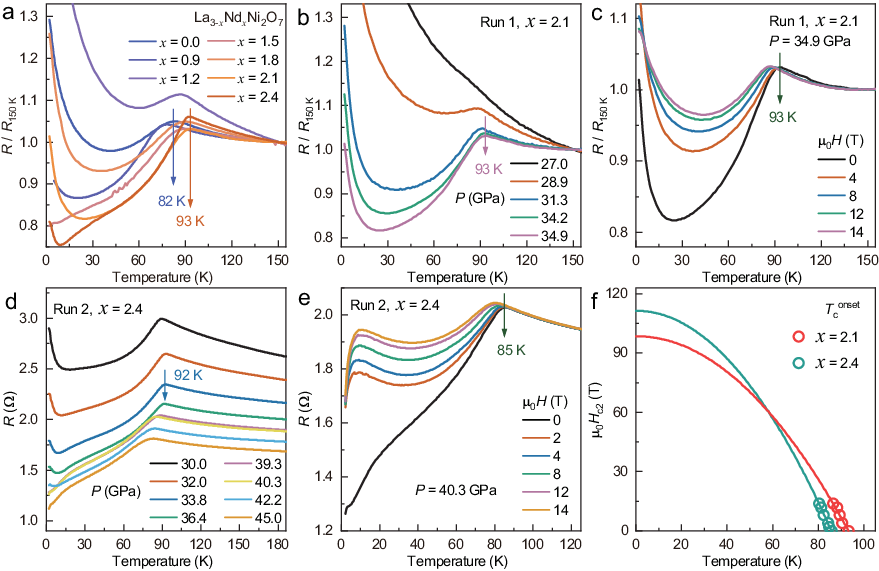}
    \caption{\textbf{ High-pressure transport properties of La$_{3-x}$Nd$_x$Ni$_2$O$_7$.}
\textbf{a,} Temperature dependence of the normalized resistance $R/R_{150\text{K}}$ for La$_{3-x}$Nd$_x$Ni$_2$O$_7$ ($0 \leq x \leq 2.4$). The blue and dark orange arrows indicate the superconducting onset temperatures $T_\text{c}^{\text{onset}}$ for the $x = 0$ and $x = 2.4$ compositions, respectively.
\textbf{b,} Normalized resistance $R/R_{150\text{K}}$ versus temperature for La$_{0.9}$Nd$_{2.1}$Ni$_2$O$_7$ at pressures from 27.0 GPa to 34.9 GPa (Run 1). The purple arrow marks the $T_\text{c}^{\text{onset}}$ of 93 K at 34.9 GPa.
\textbf{c,} Magnetic field dependence of the normalized resistance $R/R_{150\text{K}}$ for La$_{0.9}$Nd$_{2.1}$Ni$_2$O$_7$ at 34.9 GPa.
\textbf{d,} Temperature-dependent resistance of La$_{0.6}$Nd$_{2.4}$Ni$_2$O$_7$ under pressures ranging from 30.0 GPa to 45.0 GPa (Run 2). The steel blue arrow indicates the $T_\text{c}^{\text{onset}}$ of 92 K at 33.8 GPa.
\textbf{e,} Magnetic field dependence of the resistance for La$_{0.6}$Nd$_{2.4}$Ni$_2$O$_7$ at 40.3 GPa.
\textbf{f,} Upper critical fields for La$_{0.9}$Nd$_{2.1}$Ni$_2$O$_7$ at 34.9 GPa and La$_{0.6}$Nd$_{2.4}$Ni$_2$O$_7$ at 40.3 GPa. The hollow circles represent experimental $T_\text{c}^{\text{onset}}$ data, and the solid lines show fits using the Ginzburg-Landau model.}
    \label{fig2}
\end{figure*}

\subsection*{High-pressure transport properties}

To investigate the superconductivity under high pressure, we systematically measured the electronic transport properties.  As shown in Fig. \ref{fig2}a, we present the temperature-dependent resistance curves of La$_{3-x}$Nd$_x$Ni$_2$O$_7$ with different doping levels, measured under various pressures. All curves exhibit the highest $T_\text{c}$s achieved at their respective doping concentrations. As the scale of Nd doping increases, the $T_\text{c}$ rises continuously, from 82 K for the undoped sample to 93 K for the $x=2.1$ Nd-doped compound. The non-zero resistance in the low-temperature range can be attributed to the polycrystalline nature of the sample and the inhomogeneous pressure conditions resulting from the diamond anvil cell (DAC) method, which utilizes a solid pressure transmission medium (PTM)\cite{Feng2024,Zhang2024eff,Chen2025l,Zhang2024h}. Figure \ref{fig2}b displays the temperature-dependent resistance of La$_{0.9}$Nd$_{2.1}$Ni$_2$O$_7$ at various pressures. At 27.0 GPa, the sample shows semiconducting behavior with a weak anomaly in resistance. As the pressure increases slightly, a distinct transition is observed at 86 K, indicating a superconducting transition. The value of the superconducting transition pressure is approximated to the structural transition pressure of 26 GPa. It suggests an intimate correlation between structural transition and superconductivity. The $T_\text{c}$ increases with applied pressure and reaches the maximum value of 93 K at 34.9 GPa, which is the highest among all nickelate superconductors. Figure \ref{fig2}c displays the evolution of resistance curves at 34.9 GPa under various magnetic fields, ranging from 0 to 14 T, revealing the suppression of superconductivity by magnetic fields.

\begin{figure*}[t]
\centering
  \includegraphics[width=1.8\columnwidth]{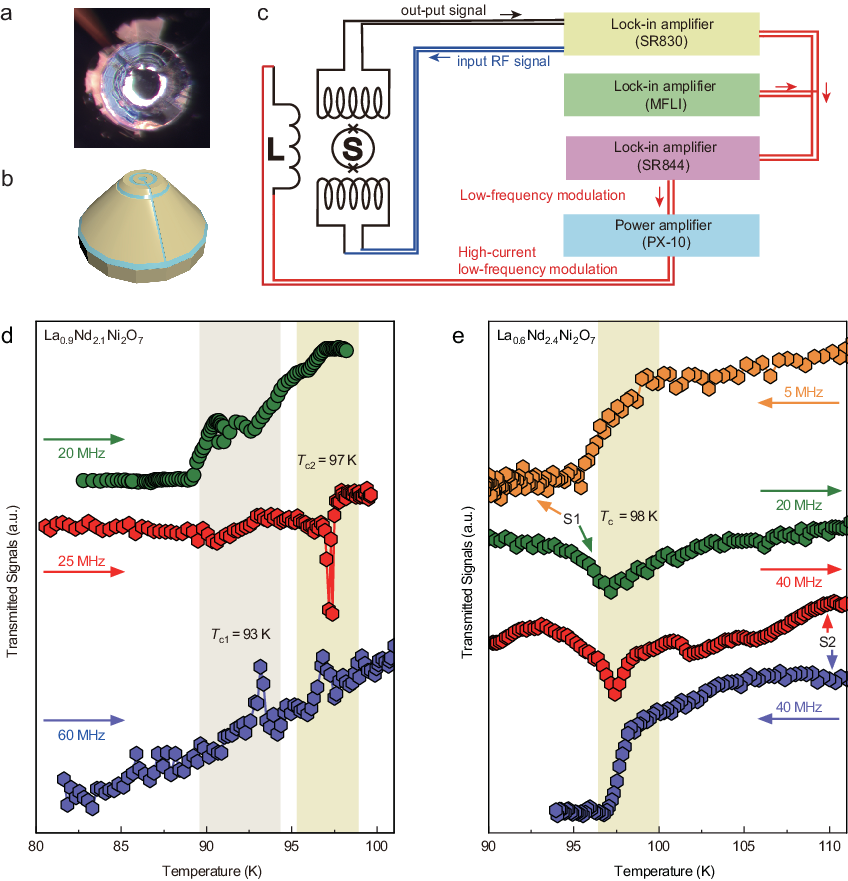}
    \caption{ \textbf{ Radio-frequency measurements of La$_{3-x}$Nd$_x$Ni$_2$O$_7$ ($x = 2.1$ and $2.4$).}
\textbf{a, b,} Optical image and front-view schematic of the RF DAC culets.
\textbf{c,} Experimental setup for RF transmission measurements in a DAC. An external solenoid (L) generates a low-frequency modulating magnetic field, which interacts with the sample (S).
\textbf{d,} Temperature dependence of the transmitted RF signal for the $x=2.1$ sample at $36 \pm 4$~GPa during warming cycles, showing the real (20~MHz) and imaginary (25 and 60~MHz) components.
\textbf{e,} Real component of the transmitted signal versus temperature for two $x=2.4$ samples (S1 and S2) at 41~GPa. For S1, data were collected at 5~MHz (cooling) and 25~MHz (warming); for S2, data were taken at 40~MHz during both cooling and warming cycles.}
    \label{fig3}
\end{figure*}

The temperature-dependent resistance of the $x=2.4$ compound is presented in Fig. \ref{fig2}d. This sample exhibits a maximum $T_\text{c}$ of 92 K. The superconductivity is further supported by the response of resistance to magnetic fields at 40.3 GPa, as shown in Fig. \ref{fig2}e. The abnormal transition at low temperatures may be attributed to the other intergrowth RP phases. Using the $T_\text{c}$s extracted from Figs. \ref{fig2}c,e, we fitted their upper critical fields to the Ginzburg-Landau formula form $\mu_0H_{c2}$($T$) = $\mu_0H_{c2}$(0)[1-($T$/$T_c)$$^2$], yielding the fitted upper critical fields of 98.5 T at 34.9 GPa for $x=2.1$ and 111.4 T at 40.3 GPa for $x=2.4$, respectively. We notice the $\mu_0H_{c2}$(0) is lowered by doping although the $T_\text{c}$ is enhanced.

\subsection*{Radio-frequency detections}

The radio-frequency (RF) transmission technique detects superconductivity by monitoring changes in magnetic permeability and surface impedance~\cite{Sakakibara1989,Timofeev2002,Semenok2025}. This method has been successfully employed to confirm high-$T_\text{c}$ superconductivity in several superhydrides at pressures above 1.5~Mbar~\cite{Semenok2025te,Semenok2025tr,Semenok2025}. 

Figure \ref{fig3}a shows an optical image of the RF DAC culet, highlighting the sample and the radial cut in the Lenz lens, while Fig.~\ref{fig3}b provides a front view. A low-frequency modulated magnetic field (Fig.~\ref{fig3}c), with an amplitude of tens to hundreds of Gauss, periodically drives parts of the superconductor into the normal state by exceeding the lower critical field ($\mu_0H_{c1}$) or the penetration field ($\mu_0H_p$). This modulation generates a second harmonic in the RF transmission signal, providing a distinct and sensitive signature of superconductivity. Prior to our measurements, we validated the method using a Bi$_2$Sr$_2$CaCu$_2$O$_{8+\delta}$ (BSCCO) film with a known $T_\text{c}$ of 94~K (Extended Data Fig.~5a).

RF transmission measurements on the $x=2.1$ compound at $36\pm4$~GPa revealed anomalies near 93~K and $97\pm2$~K at 20, 25, and 60~MHz during warming cycles (Fig.~\ref{fig3}d and Extended Data Figs.~5b--d), indicating the emergence of inhomogeneous superconductivity. Measurements on two $x=2.4$ samples followed different protocols: S1 was measured at 5~MHz (cooling) and 25~MHz (warming), while S2 was measured at 40~MHz during both cooling and warming (Fig.~\ref{fig3}e and Extended Data Figs.~5e--h). Pronounced anomalies at $98\pm2$~K confirm the systematic and reproducible nature of the superconductivity. The different responses of the transmission between cooling and warming cycles are related to the intrinsic properties of the RF method~\cite{Semenok2025}.

We note a discrepancy between the superconducting transition temperatures obtained from RF and electronic transport measurements. This may arise because the trajectory of induction currents in powder samples depends on the distribution of superconducting granules. The current path can change dramatically when some granules become superconducting, while others may have only a minor effect on the overall current distribution.

\subsection*{Temperature-pressure phase diagram}

\begin{figure}[t]
\centering
  \includegraphics[width=0.9\columnwidth]{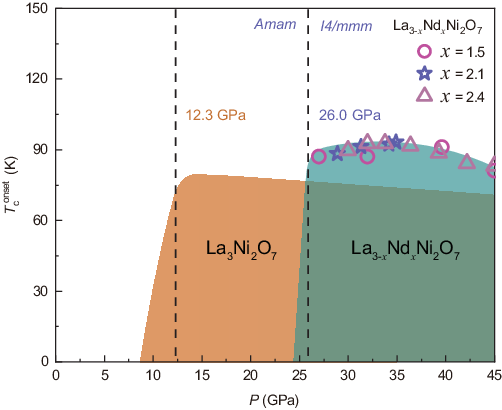}
    \caption{\textbf{ A phase diagram of La$_{3-x}$Nd$_x$Ni$_2$O$_7$}. The carmine circle, blue-gray pentacle, and purple-gray triangle represent the onset $T_\text{c}$ of the resistance curves for the $x$ = 1.5, 2.1, 2.4 compounds. Those points depict a cyan area representing the superconducting region of  $x=2.1$. The saffron yellow area symbolizing superconductivity of $x=0$, which was cited from Ref\cite{Li2025i}. The dashed lines represent structural transitions of the $x=0$ and $x=2.1$ compounds, respectively.}
            \label{fig4}
\end{figure}

Figure \ref{fig4} presents a temperature-pressure ($T$-$P$) phase diagram for heavily Nd-doped La$_{3-x}$Nd$_x$Ni$_2$O$_7$ ($x$=1.5, 2.1, 2.4), derived from high-pressure resistance and XRD measurements, compared with pristine La$_3$Ni$_2$O$_7$\cite{Li2025i}. For $x=2.1$, the structural transition from orthorhombic $Amam$ to tetragonal $I4/mmm$ occurs at $\sim$26.0 GPa, significantly higher than the 12.3 GPa required for La$_3$Ni$_2$O$_7$. The substantial increase in transition pressure with Nd doping is attributed to the enhanced orthorhombic distortions introduced by chemical substitution. These inherent distortions stabilize the low-pressure $Amam$ phase, thus raising the energy barrier for the transition to the high-symmetry $I4/mmm$ structure. With increasing pressure, the superconductivity and structural transition occur coincidentally. The $T_\text{c}$s for the Nd-heavily doped compounds are obviously increased, reaching a maximum of 93 K at 34.9 GPa for the $x=2.1$ compound. The contrast between the $T_\text{c}$s from the electronic transport and RF measurements indicates inhomogeneity of the Nd doping levels.

\subsection*{Correlations between lattice and superconductivity}

\begin{figure}[t]
\centering
  \includegraphics[width=0.9\columnwidth]{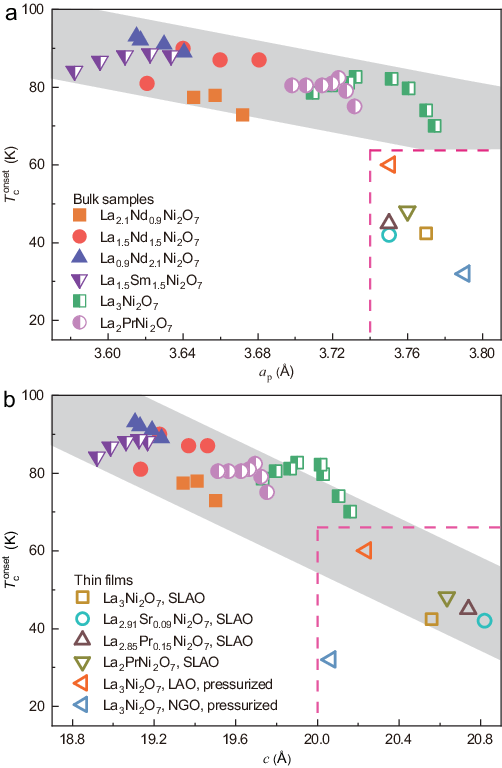}
    \caption{\textbf{ Relationship between $T_\text{c}^{\text{onset}}$ and lattice parameters.}
\textbf{a,} $T_\text{c}^{\text{onset}}$ as a function of the in-plane lattice parameter $a_\text{p}$.
\textbf{b,} $T_\text{c}^{\text{onset}}$ as a function of the out-of-plane lattice parameter $c$.
Solid symbols represent data for bulk La$_{3-x}$Nd$_x$Ni$_2$O$_7$ ($x = 0.9, 1.5, 2.1$) from this work. For comparison, open symbols show literature data for bulk La$_3$Ni$_2$O$_7$, La$_2$PrNi$_2$O$_7$, and La$_{1.5}$Sm$_{1.5}$Ni$_2$O$_7$, as well as for thin films of La$_3$Ni$_2$O$_7$ and related doped compounds under high pressure or at ambient conditions. A pink dashed box highlights data points corresponding to thin-film samples. Part of the data are adoped from Ref.\cite{Li2025i,Wang2024b,Ko2025,Hao2025,Zhou2025a,Liu2025s,Osada2025}.}
            \label{fig5}
\end{figure}

Through XRD and resistance measurements under high pressure, we correlate the onset $T_\text{c}$ of La$_{3-x}$Nd$_x$Ni$_2$O$_7$ with the in-plane parameter $a_\text{p}$ (defined as $a_\text{p}=(a+b)/2$) and the out-of-plane parameter $c$, as shown in Figs. \ref{fig5}a,b. These results are compared with reference data obtained from both pressurized bulk and thin-film samples of bilayer nickelates\cite{Li2025i,Wang2024b,Ko2025,Hao2025,Zhou2025a,Liu2025s,Osada2025}. To mitigate the influence of microcracks introduced during pressurization, we have selected the $T_\text{c}$ measured near the onset pressure and at the maximum transition temperatures. 

Factors including strain, chemical doping, and substrate-induced effects collectively modify the in-plane lattice parameter $a_\text{p}$ within a narrow scale of only $\sim$5\%. However, these factors regulate the out-of-plane parameter $c$ over a significantly wider range of $\sim$10\%. Strikingly, a clear linear relationship emerges between $T_\text{c}$ and the out-of-plane lattice constant $c$ from both pressurized bulk and thin film samples, with the exception of an over-stretched La$_3$Ni$_2$O$_7$ thin film grown on a NdGaO$_3$ substrate\cite{Li2025i,Wang2024b,Ko2025,Hao2025,Zhou2025a,Liu2025s,Osada2025}. A linear fit yields a slope of approximately $-28$ K/\AA, indicating a strong negative correlation (Fig. \ref{fig5}b). This trend can be understood within a framework where interlayer magnetic exchange coupling enhances superconductivity in these bilayer nickelates. In contrast, $T_\text{c}$ exhibits distinct and non-universal behavior as a function of the in-plane parameter $a_\text{p}$, although the value of $a_\text{p}$ appears to be a critical prerequisite for the emergence of superconductivity\cite{Ko2025,Liu2025s}.

\section*{Conclusion}

In summary, we have successfully synthesized bilayer-phase polycrystalline La$_{3-x}$Nd$_x$Ni$_2$O$_7$ ($0 \leq x \leq 2.4$) with record-high Nd-doping levels. Our comprehensive high-pressure study reveals a complex interplay between chemical doping, structural evolution, and superconductivity. We demonstrate that Nd doping enhances the orthorhombic distortion, thereby increasing the pressure required for the structural transition from the $Amam$ to the $I4/mmm$ phase and consequently shifting the emergence of superconductivity to higher pressures. Crucially, Nd doping elevates the superconducting transition temperature to a record $T_\text{c} = 93$~K in La$_{0.9}$Nd$_{2.1}$Ni$_2$O$_7$, as determined by transport measurements. This enhancement is correlated with a strengthening of the magnetic exchange coupling, evidenced by an increase in the spin density wave transition temperature. Furthermore, our radio-frequency measurements indicate the presence of superconductivity near 100~K in this system. Most significantly, we identify a universal linear relationship between $T_\text{c}$ and the out-of-plane lattice constant $c$, which underscores that enhanced interlayer coupling, achieved through lattice compression, is the fundamental mechanism promoting high-$T_\text{c}$ superconductivity in bilayer nickelates.

\section*{Methods}

\subsection*{Sample Synthesis}

La$_{3-x}$Nd$_x$Ni$_2$O$_7$ ($0 \leq x \leq 2.4$) polycrystalline samples were synthesized by the sol-gel method following previous studies\cite{Zhang1994,Wang2024p}. Starting materials of La$_2$O$_3$ (Macklin, 99.99\%), Nd$_2$O$_3$ (Macklin, 99.9\%), and Ni(NO$_3$)$_2$$\cdot$6H$_2$O (Aladdin, 99.99\%) were mixed according to the stoichiometric ratio, and dissolved in deionized water with the addition of an appropriate amount of citric acid and nitric acid. The solution was stirred at 180$^\circ$C for about 4 h until a green gel formed. The gel was then dehydrated at $200^\circ$C for 2 h, resulting in the formation of a porous black product. The fluffy black product was ground and heated overnight at $800^\circ$C to remove excess organic matter. Finally, the precursor product was ground and pressed into pellets, then heated at 1000--1100$^\circ$C for 24 h to obtain pure-phase of La$_{3-x}$Nd$_x$Ni$_2$O$_7$ polycrystalline samples.

\subsection*{Ambient-pressure characterizations}

The phase purity and crystal structure of La$_{3-x}$Nd$_x$Ni$_2$O$_7$ at ambient conditions was determined by powder X-ray diffraction (XRD) measured using a diffractometer (Rigabu) with Cu K$_{\alpha1}$ radiation ($\lambda$ = 1.5406 \AA) (Extended Data Fig. 1). Rietveld refinements of the XRD patterns were processed using the Fullprof software\cite{Rietveld1969}. The temperature-dependent electrical resistance and magnetic susceptibility were measured from 2 to 300 K at ambient pressure, as shown in Extended Data Fig. 2, using a physical properties measurement system (PPMS, Quantum Design) and a magnetic properties measurement system (MPMS, Quantum Design).

\subsection*{High-pressure structural and electrical transport measurements}

High-pressure synchrotron XRD studies were conducted at the 17UM beamline of the Shanghai Synchrotron Radiation Facility (SSRF), using an X-ray wavelength of 0.4834 \AA. The experiment utilized a symmetric DAC equipped with Boehler-Almax diamonds with a 250 $\mu$m culet. The sample was loaded into a pre-indented rhenium gasket hole, with neon as the PTM. The pressure was calibrated by the fluorescence wavelength shift of ruby spheres in the sample site. Two-dimensional diffraction patterns were collected using a flat panel detector (PILATUS R CdTe) and integrated into one-dimensional diffraction profiles using the Dioptas software (calibrated with CeO$_2$). Rietveld refinements were carried out using the FullProf software\cite{Rietveld1969}.

The experiment utilized a BeCu-type DAC (with anvil culet diameter of 300 $\mu$m) to conduct high-pressure resistance measurements. A rhenium gasket was used, insulated with a cubic boron nitride-epoxy resin mixture, and a sample chamber approximately 110 $\mu$m in diameter was pre-drilled in the insulating gasket. KBr served as the PTM to provide a quasi-hydrostatic environment. High-pressure electrical measurements were performed using the van der Pauw four-probe method. Below 30 GPa, pressure was calibrated by monitoring the fluorescence wavelength shift of ruby in the sample chamber; above 30 GPa, pressure was calibrated using the Raman spectrum of diamonds in the sample region. The two pressure calibration methods showed a discrepancy of approximately 5\% to 10\%. All electric transport measurements under high pressure were conducted on a PPMS by Quantum Design, covering a temperature range of 2 to 300 K with an applied magnetic field of up to 14 T.

\subsection*{Radio-frequency measurements}

Polycrystalline samples, pressed into pellets with a diameter of 90~$\mu$m, were loaded into a DAC equipped with diamonds having a culet of 100~$\mu$m beveled to 300~$\mu$m. Ammonia borane (NH$_3$BH$_3$) was employed as the PTM, a choice commonly used in superhydride research\cite{Sakakibara1989,Timofeev2002,Semenok2025,Semenok2025te,Semenok2025tr}.

For the radio-frequency (RF) DAC configuration, a 300-$\mu$m-thick tungsten gasket was pre-indented to 20 GPa. A sample chamber was then drilled with a diameter of 150--200\% of the diamond culet size. Both surfaces of the gasket were subsequently coated with a 1--2~$\mu$m tantalum layer via magnetron sputtering. The gasket was then heated in air above 1000~$^\circ$C to oxidize the tantalum, forming a Ta$_2$O$_5$ insulating layer. This W/Ta/Ta$_2$O$_5$ structure, with a contact resistance exceeding 100~k$\Omega$, effectively insulated the tungsten gasket from the Lenz lenses deposited on the diamond anvils.

The Lenz lenses were fabricated from deposited Ta/Au layers (total thickness 1--2~$\mu$m) using an accelerated Ar$^{+}$ beam (7.5~keV). The ion-etched zones separating the conducting rings were approximately 5~$\mu$m wide. We utilized three-stage Lenz lenses with annular etching zone diameters of 100~$\mu$m, 300~$\mu$m, and 3~mm.

The RF DAC electrical circuit incorporated an emitter consisting of a single-turn coil patterned on a 0.3 mm thick Teflon printed circuit board (PCB). The conductive patterns, made of 0.1~mm thick Cu or Au, had diameters of 3--4~mm.

Radio-frequency transmission measurements through the Lenz lens system were performed using a circuit with two or three lock-in amplifiers. The Lenz lenses were fabricated using focused ion beam (FIB) techniques with a Thermo Fisher Scientific Helios DualBeam system (Gallium ions). An SR844 lock-in amplifier (Stanford Research) served as both the high-frequency signal generator and receiver for the transmitted signal. The sample was placed in a low-frequency magnetic field (19 Hz, with a maximum induction ranging from 20 to 100 Gauss, depending on the thermal cycle) generated by a solenoid. The solenoid current was driven by an SR830 lock-in amplifier coupled with a Yamaha PX10 power amplifier, typically operating at 3--5~A. The envelope of the high-frequency signal near the superconducting transition contained even harmonics, with the strongest second harmonic detected by either an SR830 or MFLI (Zurich Instruments) lock-in amplifier.

\section*{Data availability}
Source data are provided with this paper.

\backmatter

\bmhead{Acknowledgements}

This work was supported by the National Natural Science Foundation of China (Grant No. 12425404, 12494591, 12474137), the National Key Research and Development Program of China (Grants No. 2023YFA1406000, 2023YFA1406500, 2024YFA1613100), the Guangdong Basic and Applied Basic Research Funds (Grant No. 2024B1515020040, 2025B1515020008, 2024A1515030030), the Shenzhen Science and Technology Program (Grants No. RCYX20231211090245050), the Guangzhou Basic and Applied Basic Research Funds (Grant No. 2024A04J6417), the CAS Superconducting Research Project (Grant No. SCZX-0101), the Fundamental Research Funds for the Central Universities, Sun Yat-sen University (Grant No. 74130-31610055), the Guangdong Provincial Key Laboratory of Magnetoelectric Physics and Devices (Grant No. 2022B1212010008), and the Research Center for Magnetoelectric Physics of Guangdong Province (Grant No. 2024B0303390001). V. V. S. acknowledges the support by the National Key Research and Development Program of China (Grants No. 2023YFA1608900, 2023YFA1608903). We also thank the BL17UM station and the User Experiment Assist System of the Shanghai Synchrotron Radiation Facility for help in characterization.

\section*{Author contributions}

M.W. supervised this research. Z.Y.Q., J.F.C., P.Y.M., X.H., M.W.H., T.X., and X.C. synthesized and characterized the samples. D.V.S., D.Z., H.K.M., and V.S. performed the high-pressure RF measurements. Q.Y.Z., J.Y.L., and H.L.S. conducted the high-pressure electronic transport and XRD measurements. M.W., H.L.S., D.V.S., Z.Y.Q. wrote the paper with inputs from all coauthors.

\section*{Competing interests}

The authors declare no competing interests.

\bibliography{reference}

\begin{figure*}[h]
\centering
  \includegraphics[width=1.8\columnwidth]{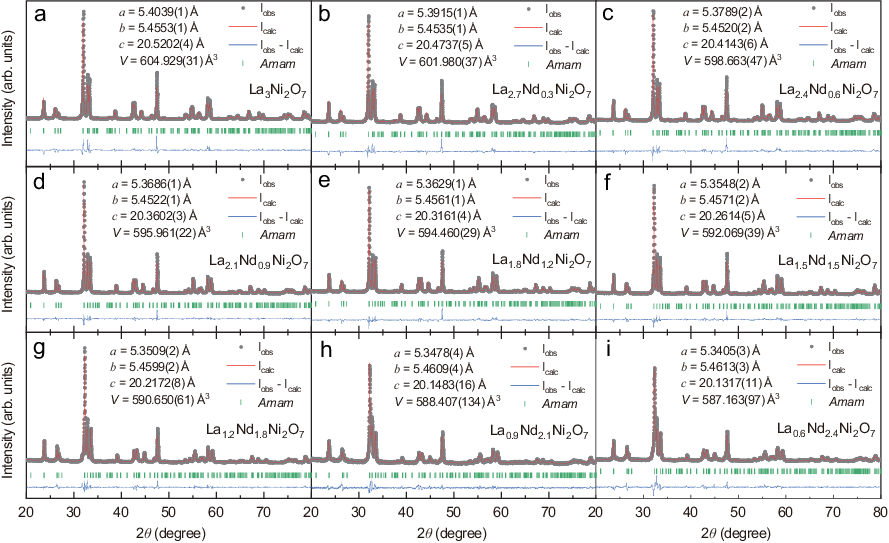}
    \caption*{ \textbf{Extended Data Fig. 1 Rietveld refinements for the La$_{3-x}$Nd$_x$Ni$_2$O$_7$ ($0 \leq  x  \leq 2.4$).} The data were collected with an X-ray diffractometer at room temperature. The nominal compositions, refined parameters, and refined diffraction patterns have been presented in the figure. }
\vspace*{6in}
\end{figure*}

\begin{figure*}[h]
\centering
  \includegraphics[width=1.8\columnwidth]{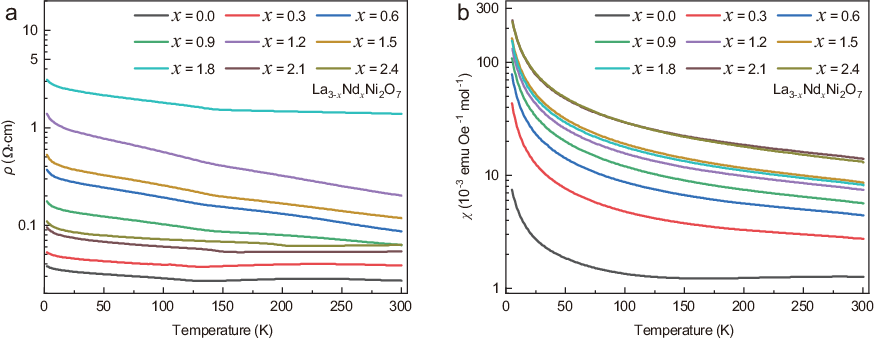}
    \caption*{ \textbf{Extended Data Fig. 2. Resistivity and magnetic susceptibility of La$_{3-x}$Nd$_x$Ni$_2$O$_7$ ($0 \leq x \leq 2.4$). a,} $\rho(T)$ of La$_{3-x}$Nd$_x$Ni$_2$O$_7$ ($0 \leq x \leq 2.4$). As Nd doping increases, the temperature dependence of $\rho(T)$ increases up to $x = 1.8$ then decreases. \textbf{b,} $\chi(T)$ of La$_{3-x}$Nd$_x$Ni$_2$O$_7$ ($0 \leq x \leq 2.4$). The temperature dependence of $\chi(T)$ increases gradually with the Nd-doping content.}
\end{figure*}

\begin{figure*}[h]
\centering
  \includegraphics[width=1.8\columnwidth]{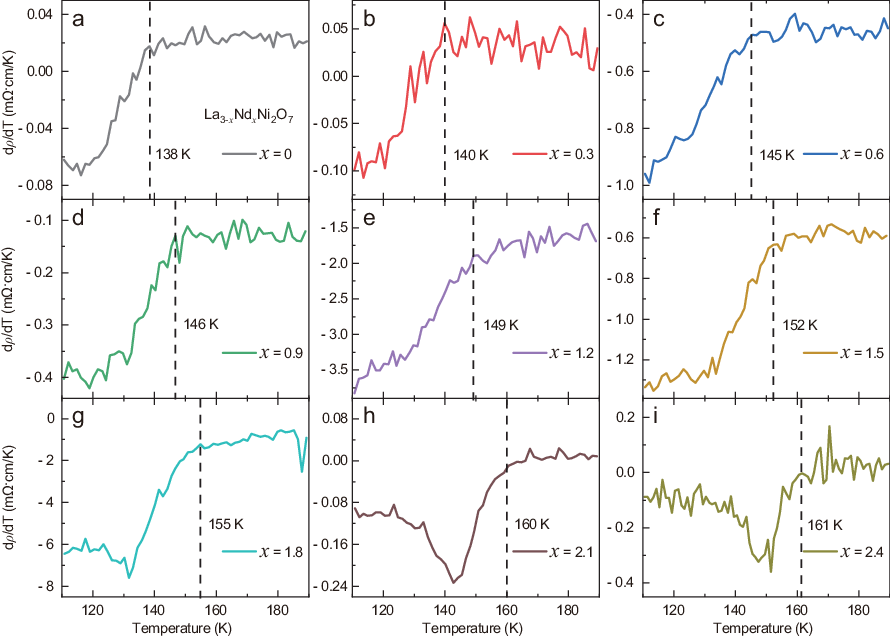}
    \caption*{ \textbf{Extended Data Fig. 3. The derivative of $\rho(T)$ against temperature of La$_{3-x}$Nd$_x$Ni$_2$O$_7$ ($0 \leq x \leq 2.4$).}  The dashed lines indicate the density wave transition temperature $T_\text{DW}$ for each composition. As Nd doping increases, the transition temperature $T_\text{DW}$ of d$\rho(T)$/d$T$ increases from 138 K to 161 K. The error of $T_\text{DW}$ is less than 3 K for each component.}
\end{figure*}

\begin{figure*}[h]
\centering
  \includegraphics[width=1.8\columnwidth]{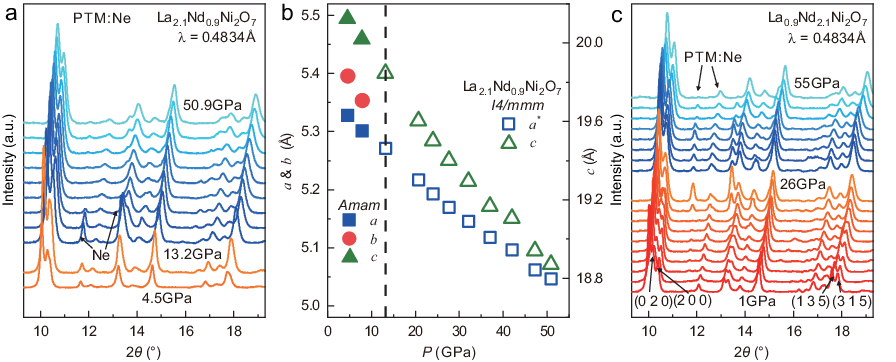}
    \caption*{ \textbf{Extended Data Fig. 4. Synchrotron XRD measurements on La$_{2.1}$Nd$_{0.9}$Ni$_2$O$_7$ and La$_{0.9}$Nd$_{2.1}$Ni$_2$O$_7$. a,} Room-temperature synchrotron XRD patterns on the $x=0.9$ compound under different pressures. The structural transition from $Amam$ to $I4/mmm$ occurs at 13.2 GPa. \textbf{b,} Lattice parameters of $a$, $b$, and $c$ refined from synchrotron XRD patterns as a function of pressure. Solid symbols represent lattice parameters in the $Amam$ space group while hollow symbols represent lattice parameters in the $I4/mmm$ space group. \textbf{c,} Identical XRD patterns on the $x=2.1$ compound under various pressures. The structural transition from $Amam$ to $I4/mmm$ occurs at 26 GPa.}
\end{figure*}

\begin{figure*}[h]
\centering
  \includegraphics[width=1.8\columnwidth]{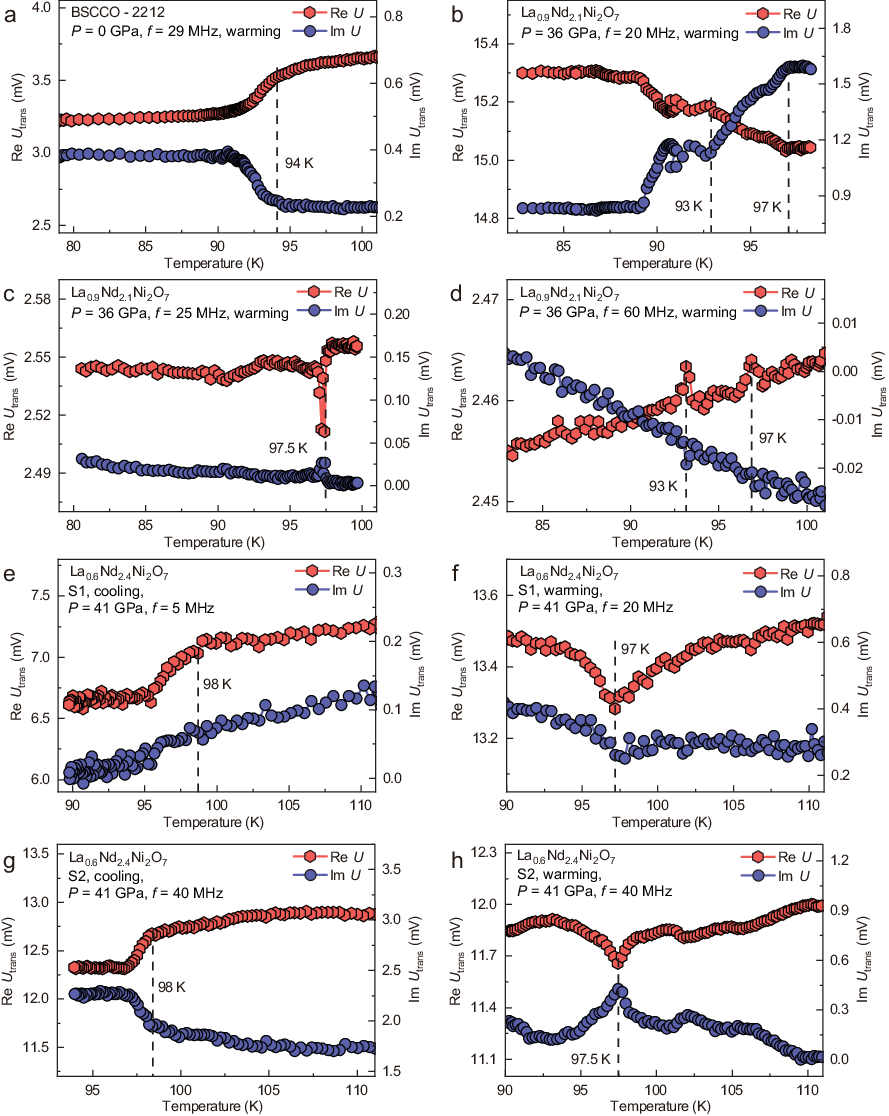}
    \caption*{\textbf{Extended Data Fig. 5. Radio-frequency transmission measurements for calibration and sample characterization.}
Real (Re $U$) and imaginary (Im $U$) components of the transmitted RF signal as a function of temperature for:
\textbf{a,} a BSCCO reference film;
\textbf{b--d,} the La$_{0.9}$Nd$_{2.1}$Ni$_2$O$_7$ sample under various conditions;
\textbf{e--h,} the La$_{0.6}$Nd$_{2.4}$Ni$_2$O$_7$ samples under different measurement protocols.}
\end{figure*}

\textbf{e} Real component of the transmitted signal as a function of temperature during a warming cycle for the same La$_{0.9}$Nd$_{2.1}$Ni$_2$O$_7$ sample at 40 GPa, measured at 25 MHz. \textbf{f,g,} Real (Re $U$) and imaginary (Im $U$) components of the transmitted RF signal as a function of temperature for the La$_{0.6}$Nd$_{2.4}$Ni$_2$O$_7$ sample at 41 GPa, measured at 40 MHz, including both a cooling cycle and a warming cycle.

\end{document}